# Method to determine argon metastable number density and plasma electron temperature from spectral emission originating from four 4p argon levels

Davide Mariotti,[a] Yoshiki Shimizu, Takeshi Sasaki, and Naoto Koshizaki
*Nanoarchitectonics Research Center (NARC), National Institute of Advanced Industrial Science and Technology (AIST), AIST Tsukuba Central 5, 1-1-1 Higashi, Tsukuba, Ibaraki 305-8565, Japan*

A simple model and method is proposed here to determine argon metastable number densities and electron temperature with the assumption of a Maxwell-Boltzmann electron energy distribution. This method is based on the availability of experimental relative emission intensities of only four argon lines that originate from any of the 4p argon levels. The proposed model has a relatively wide range of validity for laboratory plasmas that contain argon gas and can be a valuable tool for the emerging field of atmospheric microplasmas, for which diagnostics is still limited. © *2006 American Institute of Physics*. [DOI: 10.1063/1.2390631]

Plasmas are finding interest in a large number of applications and are becoming increasingly exploitable in already established plasma fields such as in material processing. This trend in plasma science is mainly due to a greater knowledge of the behavior of laboratory and technological plasmas. Diagnostic techniques play a major role in understanding plasmas and can help predicting the potential of plasmas when used for future applications. In particular, emission spectroscopy has revealed extremely useful insights when supported by very simple computational techniques.[1–7] One of the advantages of emission spectroscopy is undoubtedly the noninvasive character; the practicality and the availability of low cost spectroscopy equipment have also been an incitement to the use of this diagnostic technique.

Probably one of the most important plasma parameter is represented by the electron energy distribution function (EEDF). The EEDF can be easily described by a single parameter, the electron temperature ($T_e$), when electrons assume a Maxwell-Boltzmann energy distribution (or not far from Maxwell-Boltzmann behavior). If the actual energy distribution of electrons strongly deviates from a Maxwell-Boltzmann behavior, the use of an *effective* or *equivalent electron temperature* is also useful and commonly accepted. The interpretation of an equivalent electron temperature ($T_{eq}$) is sometimes given by the formula $T_{eq}=(2/3)\varepsilon_{av}$, where $\varepsilon_{av}$ is the average electron energy of the true EEDF.[2] Alternatively, an effective electron temperature is defined as the temperature required by a Maxwell-Boltzmann EEDF to satisfactorily model the behavior of plasma with respect to specific experimental evidence.[5,8–10]

Argon and other rare gases are very common plasma constituents in many applications and are also used in fundamental plasma studies. Argon, for instance, can be either a main gas component or it can be added to the plasma gas mixture with very little perturbation. In both cases, argon can be used to "spectroscopically probe" the plasma, providing useful diagnostics such as information on the electron temperature.[6] In argon-based plasmas, another important parameter is the number density of argon metastable levels.[11] Metastable levels play an important role in the energy kinetics, as they are *de facto* energy reserves.[12]

In this work, we show how a very simple collisional-radiative model can be used to determine at the same time the electron temperature and number densities of metastable levels by knowing relative emission intensities of only four argon lines. The assumptions made for this model, to be valid, are reasonable for a wide range of plasmas used in current technology applications.

In these calculations a Maxwell-Boltzmann distribution for the EEDF is assumed, and experimental emission data are utilized so that an effective electron temperature is obtained. Techniques to overcome this limitation are under development, whereby no knowledge of the EEDF is required *a priori*. The model is based on the balance equation of the excited level density of four of any of the ten 4p levels. The model is self-consistent in the sense that does not require assumptions or knowledge of any of the other energy levels.

The validity of the model is limited to plasma where electron excitation kinetics is dominant, and the application of spectral emission in plasma diagnostics requires the plasma to be optically thin. In principle, the use of plasma diagnostics by emission spectroscopy also requires that the acquisition of the emission spectra should be performed over a period of time, in which the plasma parameters can be considered sufficiently constant and uniform over the volume where plasma emission originates. Finally, electron density should be sufficiently low to consider radiative decay, the main depopulation process, similarly to corona models.[8]

The assumptions above are quite general and valid for many low-pressure plasmas, less than 1 Torr in optically thin conditions.[4,6] At higher argon partial pressure, processes involving neutrals and ions may become important.[9,13] Nevertheless if the plasma remains a nonequilibrium plasma, electron excitation kinetics still plays a major role over other processes;[14] for instance, this can be true for atmospheric pressure microplasmas.[15,16] In this case, the range of validity is determined by the electron density and the rate coefficients of atom-induced excitation versus those of electron-induced excitation. Atmospheric microplasmas have shown to have electron density in the range from about $10^{21}$ m$^{-3}$ to about $10^{23}$ m$^{-3}$, and at these conditions atom-atom processes are negligible for an electron temperature above 0.7 eV.[14] The second problem arising from higher argon partial pressure is radiation trapping, in which case escape factor parameters or pressure dependent cross sections should be introduced in this model, similarly to any collisional-radiative model.[9] Microplasmas are again an exception as they have the advantage of maintaining a high ratio between surface area and plasma depth, so that in some cases can still be considered optically thin at higher pressure ranges.[9] Another possibility to avoid the effects of radiation trapping is to carefully choose emission lines that are little affected by reabsorption phenomena. As it will be shown below, only four emission lines are necessary for the calculations and a relatively wide choice is therefore available.

---

[a]Electronic mail: d.mariotti@aist.go.jp

Accordingly to the above assumptions, population of the $4p$ excited argon levels ($2p_1$–$2p_{10}$ in Paschen notation) mainly occurs by electron collision with argon atoms in the ground level and with atoms in the metastable levels ($1s_3$ and $1s_5$ in Paschen notation). Cascading from higher excited states ($3d$, $5s$, $4d$, and $6s$) also contributes to the population of the $4p$ levels. Depopulation processes from the $4p$ ($2p_1$–$2p_{10}$ in Paschen notation) levels are represented by radiative decay to the $4s$ states ($1s_2$–$1s_5$ in Paschen notation). The balance equation for a $2p_x$ (Paschen notation) excited level can therefore be written as

$$n_e n_g k_{g,2p_x} + \sum_{i=3,5} n_e n_{1s_i} k_{1s_i,2p_x} + \sum_i n_{\text{higher},i} A_{i,2p_x}$$

$$= \sum_{i=5}^{2} n_{2p_x} A_{2p_x,1s_i}, \quad (1)$$

where $n$ indicates number density, $k_{x,y}$ is the rate coefficient for the excitation from level $x$ to $y$, and $A_{y,x}$ is the transition probability from level $y$ to $x$. The subscript $g$ stands for ground level, $e$ for electrons, and $1s_x$ and $2p_x$ are the Paschen notations for each $3p$ and $4p$ energy levels respectively.

In Eq. (1), the cascading contribution of the high energy levels can be ignored if sufficiently small (third term on the left side of the equation).[4] Alternatively, it is often possible to include this effect by using apparent cross sections in the calculation of the reaction rate coefficients $k_{g2p_x}$ and $k_{1s_i,2p_x}$. In the following, the cascading term is ignored assuming that one of the two solutions just mentioned is possible.

Spectral emission intensities can easily produce a set of relative number densities for several energy levels. This can be done taking into consideration the following relationship:

$$n_x = C \frac{I_{x,y} \lambda_{x,y}}{A_{x,y}} = C n'_x, \quad (2)$$

where $x$ and $y$ represent, respectively, upper and lower energy levels, $I$ is the measured relative intensity, and $\lambda$ the corresponding wavelength. The value for $C$ is the proportionality constant which relates the actual number density ($n_x$) and the relative number density ($n'_x$) available from experimental spectral data.

Wavelengths for transitions originating from the $4p$ levels range from about 650 to 1150 nm. Some examples are represented by some of the lines from levels $2p_2$ (696.543, 727.293, and 826.452 nm), $2p_3$ (706.722 and 738.398 nm), $2p_6$ (763.510, and 922.450 nm), and $2p_{10}$ (912.297 and 965.779 nm), where levels are indicated in Paschen notation.

Equation (2) can be used in Eq. (1) to replace $n_{2p_x}$ and then Eq. (1) can be rearranged as follows:

$$n'_e = \frac{n_e}{C} = \frac{\sum_{i=5}^{2} n'_{2p_x} A_{2p_x,1s_i}}{n_g k_{g,2p_x} + \sum_{i=3,5} n_{1s_i} k_{1s_i,2p_x}}. \quad (3)$$

Equation (3) can be written for any $2p_x$ argon level. The transition probability coefficients $A$ in Eq. (3) are readily available. The ground state density can be estimated with $n_g = p/k_B T$, where $k_B$ is the Boltzmann constant, $p$ is the argon partial pressure, and $T$ the gas temperature.[6] Both pressure and gas temperature must be known and can be easily measured and/or derived. It has to be underlined that the error committed in the ground state estimation is quite often irrelevant due to the dominance of excitation from metastable levels. Therefore an accurate measurement of pressure and temperature is seldom needed for these calculations. In Eq. (3), also the rate coefficients $k$ are available in some cases. Nevertheless, we will suppose here that such rate coefficients are not known, but we assume a Maxwell-Boltzmann EEDF for the electrons from which the rates can be calculated according to the following:

$$k(T_e) = \int_0^\infty \sigma(\varepsilon) \sqrt{\frac{2\varepsilon}{m}} f_M(\varepsilon, T_e) d\varepsilon, \quad (4)$$

where $\varepsilon$ is the energy, $m$ is the electron mass, and $f_M$ is the normalized Maxwell-Boltzmann EEDF. The symbol $\sigma$ stands for the cross section of the collision between electrons and the corresponding particle for the rate that is being calculated.

We can now write Eq. (3) for any of four different $2p_x$ levels (indicated as $2p_a$, $2p_b$, $2p_c$, and $2p_d$), and couple such equations two by two,

$$\frac{\sum_{i=5}^{2} n'_{2p_a} A_{2p_a,1s_i}}{n_g k_{g,2p_a}(T_e) + \sum_{i=3,5} n_{1s_i} k_{1s_i,2p_a}(T_e)}$$

$$= \frac{\sum_{i=5}^{2} n'_{2p_b} A_{2p_b,1s_i}}{n_g k_{g,2p_b}(T_e) + \sum_{i=3,5} n_{1s_i} k_{1s_i,2p_b}(T_e)},$$

$$\frac{\sum_{i=5}^{2} n'_{2p_c} A_{2p_c,1s_i}}{n_g k_{g,2p_c}(T_e) + \sum_{i=3,5} n_{1s_i} k_{1s_i,2p_c}(T_e)}$$

$$= \frac{\sum_{i=5}^{2} n'_{2p_d} A_{2p_d,1s_i}}{n_g k_{g,2p_d}(T_e) + \sum_{i=3,5} n_{1s_i} k_{1s_i,2p_d}(T_e)}. \quad (5)$$

Equations (5) can then be solved with respect to metastable number densities $n_{1s_3}$ and $n_{1s_5}$. The analytical solution for the metastable densities will be of the form shown below,

$$n_{1s_3} = n_{1s_3}(T_e),$$

$$n_{1s_5} = n_{1s_5}(T_e). \tag{6}$$

The actual formulas are easily obtained with simple mathematical passages, which are omitted here for simplicity. The dependency on the electron temperature originates from the rate coefficients and Eq. (4). The solutions in Eqs. (6) have been reached by coupling Eq. (3) written for $2p_a$ with the same equation written for $2p_b$ and analogously, equation for $2p_c$ was coupled with the one written for $2p_d$. We can now write relative electron number density $n'_e$ using Eq. (3) for $2p_a$ and the same can be done for $2p_c$ using metastable number density in Eq. (6). We therefore obtain two values for the relative electron number densities, $n'_{a,e}$ and $n'_{c,e}$,

$$n'_{a,e}(T_e) = \frac{\sum_{i=5}^{2} n'_{2p_a} A_{2p_a,1s_i}}{n_g k_{g,2p_a}(T_e) + \sum_{i=3,5} n_{1s_i}(T_e) k_{1s_i,2p_a}(T_e)},$$

$$n'_{c,e}(T_e) = \frac{\sum_{i=5}^{2} n'_{2p_c} A_{2p_c,1s_i}}{n_g k_{g,2p_c}(T_e) + \sum_{i=3,5} n_{1s_i}(T_e) k_{1s_i,2p_c}(T_e)}. \tag{7}$$

The electron temperature $T_e$ can now be used as a parameter to equal $n'_{a,e}$ with $n'_{c,e}$. The procedure will provide at the same time relative electron number density ($n'_e = n'_{a,e} = n'_{c,e}$) and a value for the electron temperature. The electron temperature value can then be used in Eq. (6) to determine absolute values of the number density of the two metastable levels.

The method outlined above can therefore provide metastable number density and electron temperature. It relies on the availability of relative spectral emission intensities of only four argon lines originating from four different $4p$ levels. As mentioned above, the selection of the four argon lines can help avoiding problems arising for radiation trapping. Knowledge of the cross sections from ground levels to the four $4p$ levels by electron collision are also necessary, and can be found in the literature.[17–20] Finally, excitation cross sections by electron collision from the metastable levels to the four $4p$ levels are also required. Due to a general increasing interest in metastable states in rare gases, also these cross sections have become more available.[12,19,21,22]

The method relies on the accuracy of the cross section data, which are expected to improve in the years to come. Also, the resolution of the numerical evaluation of Eq. (4) plays an important role. Application of this method to a microplasma using theoretical cross sections from Ref. 21 has provided initial results, which will be confirmed and then reported separately. In addition, it can be noticed that the same technique can be adapted to other rare gases that share with argon a similar configuration structure.

In conclusion a very simple method has been described to determine the metastable number density from very few experimental parameters, which also provide at the same time the value for an effective electron temperature.